\documentclass{article} 
\usepackage{template/iclr2025_conference,times}
\usepackage{amsmath,amsfonts,bm}

\def\1{\bm{1}}

\def\vtau{{\bm{\tau}}}
\def\va{{\bm{a}}}

\def\vq{{\bm{q}}}

\def\vx{{\bm{x}}}
\def\vy{{\bm{y}}}

\def\mW{{\bm{W}}}

\DeclareMathAlphabet{\mathsfit}{\encodingdefault}{\sfdefault}{m}{sl}
\SetMathAlphabet{\mathsfit}{bold}{\encodingdefault}{\sfdefault}{bx}{n}

\usepackage[section,above,below]{placeins}  
\usepackage{caption}                        
\DeclareCaptionLabelSeparator{bar}{ $|$ }
\usepackage{hyperref}                       
\usepackage{url}                            
\usepackage{chngcntr}                       
\definecolor{blue}{HTML}{0975C1}
\hypersetup{
  colorlinks=true,
  citecolor=blue,
  linkcolor=blue,
  urlcolor=blue,
}

\newcommand{\figref}[2][]{\hyperref[#2]{\figureautorefname~\ref*{#2}#1}}  
\usepackage{etoolbox}
\makeatletter
\patchcmd{\hyper@makecurrent}{%
    \ifx\Hy@param\Hy@chapterstring
        \let\Hy@param\Hy@chapapp
    \fi
}{%
    \iftoggle{inappendix}{
        \@checkappendixparam{chapter}%
        \@checkappendixparam{section}%
        \@checkappendixparam{subsection}%
        \@checkappendixparam{subsubsection}%
        \@checkappendixparam{paragraph}%
        \@checkappendixparam{subparagraph}%
    }{}%
}{}{\errmessage{failed to patch}}

\newcommand*{\@checkappendixparam}[1]{%
    \def\@checkappendixparamtmp{#1}%
    \ifx\Hy@param\@checkappendixparamtmp
        \let\Hy@param\Hy@appendixstring
    \fi
}
\makeatletter

\newtoggle{inappendix}
\togglefalse{inappendix}

\apptocmd{\appendix}{\toggletrue{inappendix}}{}{\errmessage{failed to patch}}

\usepackage[utf8]{inputenc} 
\usepackage[T1]{fontenc}    
\usepackage{booktabs}       
\usepackage{amsmath}        
\usepackage{amsfonts}       
\usepackage{nicefrac}       
\usepackage{microtype}      
\usepackage{xcolor}         
\usepackage{graphicx}       
\usepackage{lipsum}         
\usepackage{wrapfig}        
\usepackage{makecell}       
\usepackage{enumitem}       

\title{Neural Circuit Architectural Priors\\for Quadruped Locomotion}

\author{%
  Nikhil~X.~Bhattasali,\enspace
  Venkatesh~Pattabiraman,\enspace
  Lerrel~Pinto,\enspace
  Grace~W.~Lindsay\\
  New York University\\
  \texttt{nikhilxb@nyu.edu}\vspace{-1pt}
}

\newcommand{\website}{\url{https://ncap-quadruped.github.io/}}
\newcommand{\webref}[1]{\href{https://ncap-quadruped.github.io/}{#1}}
\newcommand{\oscref}{\citep[in submission]{Bhattasali2024SimplifiedModelIntrinsically}}
\iclrfinalcopy

\begin{document}

\maketitle

\begin{abstract}
Learning-based approaches to quadruped locomotion commonly adopt generic policy architectures like fully connected MLPs. As such architectures contain few inductive biases, it is common in practice to incorporate priors in the form of rewards, training curricula, imitation data, or trajectory generators. In nature, animals are born with priors in the form of their nervous system's architecture, which has been shaped by evolution to confer innate ability and efficient learning. For instance, a horse can walk within hours of birth and can quickly improve with practice. Such architectural priors can also be useful in ANN architectures for AI. In this work, we explore the advantages of a biologically inspired ANN architecture for quadruped locomotion based on neural circuits in the limbs and spinal cord of mammals. Our architecture achieves good initial performance and comparable final performance to MLPs, while using less data and orders of magnitude fewer parameters. Our architecture also exhibits better generalization to task variations, even admitting deployment on a physical robot without standard sim-to-real methods. This work shows that neural circuits can provide valuable architectural priors for locomotion and encourages future work in other sensorimotor skills.
\end{abstract}

\section{Introduction}
\label{sec:1_introduction}

Learning-based approaches to quadruped locomotion commonly adopt generic policy architectures like fully connected multilayered perceptrons (MLPs) \citep{Rudin2022LearningWalkMinutes,Smith2022WalkParkLearning,Agarwal2022LeggedLocomotionChallenging}.
As such architectures contain few inductive biases, they must rely on training to develop desired behaviors. Simple reward functions often do not lead to naturalistic or robust behavior \citep{Heess2017EmergenceLocomotionBehaviours}. Therefore, it is common in practice to incorporate priors in the form of rewards \citep{Rudin2022LearningWalkMinutes}, training curricula \citep{Agarwal2022LeggedLocomotionChallenging,Rudin2022LearningWalkMinutes}, imitation data \citep{BinPeng2020LearningAgileRobotic,Merel2019NeuralProbabilisticMotor}, or trajectory generators \citep{Schaal2006DynamicMovementPrimitives,Iscen2019PoliciesModulatingTrajectory}.

In nature, animals are born with priors in the form of their nervous system's architecture, which has been shaped by evolution to confer innate ability and efficient learning \citep{Zador2019CritiquePureLearning,Cisek2019ResynthesizingBehaviorPhylogenetic}. For instance, a horse can walk within hours of birth and can quickly improve with practice, and humans have strong inductive biases for perceiving and interacting with the world \citep{Lake2017BuildingMachinesThat}. These inductive biases are a reflection of highly structured neural circuit connectivity \citep{Luo2021ArchitecturesNeuronalCircuits}, which combines innate and learning mechanisms in stark contrast to generic ANN architectures.

Can such architectural priors also be useful in ANN architectures for AI? \cite{Bhattasali2022NeuralCircuitArchitectural} investigated this by introducing Neural Circuit Architectural Priors (NCAP). Using a case study of the nematode \textit{C.~elegans}, the proposed Swimmer NCAP translated neural circuits for swimming into an ANN architecture controlling a simulated agent from an AI benchmark \citep{Tassa2020DmControlSoftware}. Swimmer NCAP achieved good performance, data efficiency, and parameter efficiency compared to MLPs, and its modularity facilitated interpretation and transfer to new body designs. As such, Swimmer NCAP demonstrated several possible advantages of biologically inspired architectural priors for AI.

However, it remained unknown whether the approach could scale to more complex animals and tasks. \textit{C.~elegans} has a nervous system of only 302 neurons and highly stereotyped connectivity, and its connectome has long been mapped \citep{White1986StructureNervousSystem}. In contrast, mammalian nervous systems have millions or billions of neurons \citep{Herculano-Houzel2006CellularScalingRules, Herculano-Houzel2007CellularScalingRules} with more variable connectivity and no mapped connectome, so it was not obvious how such circuits could inspire AI.

In this work, we introduce Quadruped NCAP, a biologically inspired ANN architecture for quadruped locomotion based on neural circuits in the limbs and spinal cord of mammals. Our key insights for scaling up the NCAP approach are: modeling at the level of genetically defined neural populations \citep{Danner2017ComputationalModelingSpinal,Ausborn2019ComputationalModelingBrainstem,Kim2022ContributionAfferentFeedback} rather than at the level of single neurons, leveraging machine learning to compensate for gaps in detailed circuit knowledge, and introducing methodological innovations to improve the architecture's expressivity and trainability (\autoref{sec:2_related_work}). Together, these insights enable the architecture to successfully control quadruped locomotion, which is a much harder task than planar swimming due to its higher dimensionality and inherent instability.

We test the value of our architectural prior by comparing to an architecture without priors: the MLP, which is commonly used in robotics. Quadruped NCAP achieves good initial (untrained) performance and comparable final (asymptotic) performance to MLPs, while using less data and orders of magnitude fewer parameters. Our architecture also exhibits better generalization to task variations, even admitting deployment on a physical robot without standard domain randomization methods that are often needed for sim-to-real generalization. This work shows that neural circuits can provide valuable architectural priors for locomotion in more complex animals and encourages future work in yet more complex sensorimotor skills.
    
The key contributions of this work are:

\begin{enumerate}[leftmargin=0.25in, nosep, parsep=1ex, itemsep=0ex]
    \item \textbf{First neural circuit model for quadrupedal robot locomotion.} We design an architecture that scales up NCAP to quadruped locomotion. While many related works in locomotion have been inspired by biology  (\autoref{sec:2_related_work}), to the best of our knowledge this work is the first to use genetically defined neural circuits to control locomotion in a standard quadruped robot.
    \item \textbf{Extensive evaluation in simulation.} We extensively evaluate NCAP in terms of performance, data efficiency, parameter efficiency, and generalization to terrain and body variations. NCAP learns more naturalistic gaits, with up to millions of fewer timesteps and orders of magnitude fewer parameters than MLP, while being more robust to unseen conditions.
    \item \textbf{Deployment on the physical robot.} We deploy NCAP to the physical robot to test its generalization across a large sim-to-real domain gap. While MLP falls immediately, NCAP walks successfully.
\end{enumerate}

Our open-source code and videos are available at: \website
\section{Related Work}
\label{sec:2_related_work}

Our work builds on an extensive literature in neuroscience, robotics, and artificial neural networks. For conciseness, we highlight the most relevant ones to our work:

\paragraph{Central Pattern Generators}
Central pattern generators (CPGs) are neural circuits that produce rhythmic activity in the absence of rhythmic inputs, and they underlie many movements including chewing, breathing, and locomotion. Roboticists have developed CPG-like controllers for a variety of tasks, and these controllers come in diverse forms \citep{Ijspeert2008CentralPatternGenerators,Yu2014SurveyCPGinspiredControl}. For example, the desired movement can be directly engineered using trajectory generators (for instance, swing/stance trajectories) that are adjusted by a higher-level controller \citep{Iscen2019PoliciesModulatingTrajectory}. Alternatively, a policy can adopt an action space of controllable abstract oscillators to flexibly modulate \citep{Bellegarda2022CPGRLLearningCentral, Shafiee2023PuppeteerMarionetteLearning}. In this work, we take a biologically constrained approach that instantiates a CPG using a network of neurons, some of which have intrinsic bursting dynamics.

\paragraph{Neuromechanical Models}
Neuromechanical models are used in computational neuroscience to develop insights about the interactions between the musckuloskeletal system and the nervous system \citep{Ausborn2021ComputationalModelingSpinal, Markin2016NeuromechanicalModelSpinal}. Recently, several neuromechanical models have been developed for the rodent \citep{Merel2019DeepNeuroethologyVirtual, TataRamalingasetty2021WholebodyMusculoskeletalModel} and the fly \citep{Lobato-Rios2022NeuroMechFlyNeuromechanicalModel, Wang-Chen2023NeuroMechFly20Framework}, which will enable new understanding about how animals perform movement. In this work, we build upon insights gleaned from neuromechanical models, but our goal is not to control a realistic musckuloskeletal simulation. Rather, we aim to translate insights from biology to AI and robotics, which leads us to model at a higher level of abstraction.

\paragraph{Architectural Priors}
Architectural priors incorporate structure into an ANN to improve performance and efficiency. For example, convolutional neural networks inspired by the visual system incorporate translation invariance 
over images \citep{Lindsay2021ConvolutionalNeuralNetworks}. In locomotion, various kinds of architectural prior have been explored, including priors on task and body symmetries \citep{Mittal2024SymmetryConsiderationsLearning,Ding2024BreakingSymmetriesLeads}, spring-loaded inverted pendulum dynamics \citep{Ordonez-Apraez2022AdaptableApproachLearn}, and hierarchy \citep{Heess2016LearningTransferModulated, Heess2017EmergenceLocomotionBehaviours}. In this work, we explore priors based on genetically defined neural circuits, similar to connnectome-constrained networks explored in the vision literature \citep{Lappalainen2024ConnectomeconstrainedNetworksPredict}. Our prior is encoded through sparse and structured connectivity, weight sign constraints, weight initializations, bilateral symmetry, and intrinsic neural dynamics.

\paragraph{Swimmer NCAP}
In this work, we build on Swimmer NCAP \citep{Bhattasali2022NeuralCircuitArchitectural} by scaling it to more complex settings. The animals we study are more complex: mammals have much larger and less understood nervous systems than nematodes. The AI tasks we tackle are also much harder: the Quadruped is a higher-dimensional inherently unstable body, while the Swimmer is a lower-dimensional inherently stable (planar) body. We also evaluate on a physical robot, in contrast to previous work in simulation only. Thus, our work addresses problems that are especially relevant for AI. Scaling to more complex settings requires novel methodological innovations: (1) We adopt a continuous-time formulation of ANNs, in contrast to the discrete-time Swimmer. (2) We build a recurrent architecture in the Rhythm Generation (RG) module, in contrast to the feedforward Swimmer. (3) We develop a closed-loop Oscillator unit that admits period modulation, phase shifts, and entrainment, in contrast to the open-loop Oscillator unit in Swimmer that does not respond to input signals. (4) We model circuits at the level of neural populations, rather than at the level of single neurons. (5) We enable the architecture to learn unknown connections in the Afferent Feedback (AF) and Pattern Formation (PF) modules by leaving certain connectivity and signs unconstrained, in contrast to the Swimmer which fully constrained connectivity and signs. Together, these novel insights and methods can help the community advance biologically inspired architectural priors.
\section{Methods}
\label{sec:3_methods}

We translate neuroscientific models of quadruped locomotion circuits into an ANN architecture for controlling a robot.
In \autoref{sec:3_biological_locomotion}, we review key background about the neuroscience of locomotion.
In \autoref{sec:3_architecture_units}, we describe the computational units that are building blocks in our Quadruped NCAP architecture.
In \autoref{sec:3_architecture_structure}, we describe the connectivity of NCAP and its interface with the robot.

\subsection{Biological Locomotion}
\label{sec:3_biological_locomotion}

\begin{figure}[t]
\centering
\includegraphics[width=\linewidth]{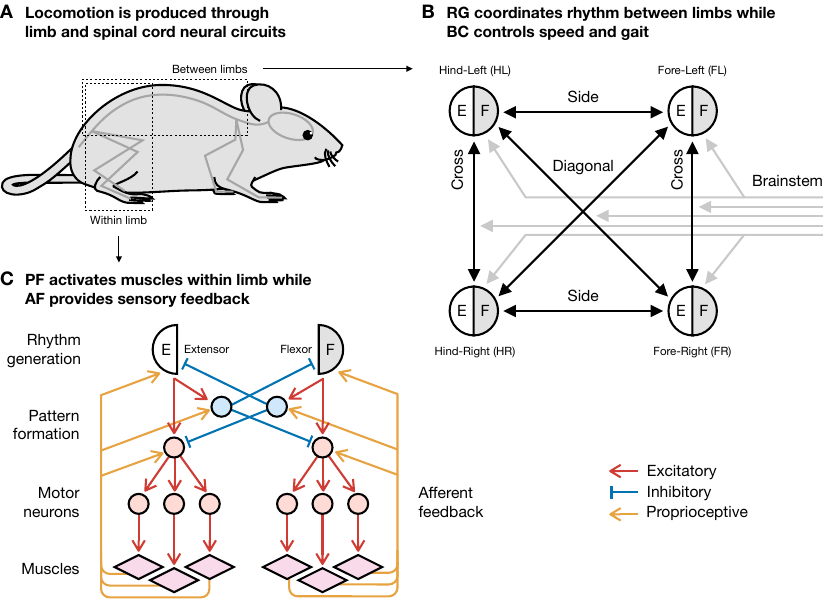}
\caption{
    \textbf{Biological Locomotion.}
    \textbf{A}, Quadruped mammals share a homologous organization of their musculoskeletal systems and neural circuits. Surprisingly, neural circuits in the limbs and spinal cord are sufficient to produce locomotion, while higher brain regions are important to initiate and regulate locomotion.
    \textbf{B}, Neural circuits for rhythm generation (RG) and brainstem command (BC), adapted from \cite{Danner2017ComputationalModelingSpinal}. Each limb is controlled by flexor (F) and extensor (E) half-centers. Between limbs, half-centers communicate through connections that promote synchonization or alternation. Brainstem command signals modulate the half-centers and connection activations.
    \textbf{C}, Neural circuits for pattern formation (PF) and afferent feedback (AF), adapted from \cite{Kim2022ContributionAfferentFeedback}. Within limbs, interneurons and motorneurons convert half-center states into specific muscle commands, while sensory feedback modulates the circuit at multiple levels. 
}
\label{fig:3_biological_locomotion}
\end{figure}

Quadrupeds locomote by rhythmically flexing and extending their limbs in a coordinated gait to propel the body forward.
They control their velocity by producing various gaits (such as walk, trot, gallop, and bound), and they adapt to different conditions using sensory information.
Quadruped mammals (including mice, cats, dogs, and horses) exhibit considerable differences in appearance, but comparative anatomical studies have revealed a remarkable homology in body structure and neural circuitry between them, which makes sense given their shared evolutionary heritage \citep{Grillner2020CurrentPrinciplesMotor}. Neuroscience research across many animal systems has shed light on how locomotion is achieved through the complex interaction between the musculoskeletal system and neural circuits in the limbs, spinal cord, and higher brain regions \citep{Grillner2020CurrentPrinciplesMotor}. Surprisingly, neural circuits in the limbs and spinal cord are sufficient to produce locomotion, while higher brain regions are important to initiate and regulate locomotion \citep{Rybak2015OrganizationMammalianLocomotor}. Classic studies strikingly demonstrated evidence of such organization using decerebrate animals, in which most of the brain was severed from the spinal cord, yet the animal could still walk and even transition between gaits when tugged along a treadmill \citep{Whelan1996ControlLocomotionDecerebrate}. Recent studies have leveraged advances in experimental tools like molecular genetics to precisely map and manipulate locomotor circuits with cell-type specificity \citep{Kiehn2016DecodingOrganizationSpinal,Ausborn2021ComputationalModelingSpinal}.

We summarize below a well-supported neuroscientific model of locomotor circuits (\figref{fig:3_biological_locomotion}), which is based on data from cats and transgenic mice, and which adopts the abstraction of genetically defined neural populations with rate-coded activity. For details, please refer to the referenced works.

\paragraph{Rhythm Generation} The RG neural circuit within the spinal cord coordinates limbs to produce gait rhythms (\figref[B]{fig:3_biological_locomotion}). Each limb is controlled by a half-center microcircuit consisting of a flexor center with intrinsically bursting neurons and an extensor center with tonic firing neurons. The paired centers inhibit each other, leading to oscillating flexion and extension in each limb. The half-centers for the four limbs communicate through ``cross'' connections (cervical/lumbar commissural interneurons), ``side'' connections (homolateral long propriospinal neurons), and ``diagonal'' connections (diagonal long propriospinal neurons), thus enabling bilateral and ascending/descending communication. In essence, excitatory connections between half-centers promote synchronization and inhibitory connections promote alternation, so the speed-dependent activation of these various connections change the timing between limbs and therefore the gait. \citep{Danner2017ComputationalModelingSpinal} 

\paragraph{Brainstem Command} The BC neural circuit in the brainstem conveys command signals to adjust locomotor speed and gait (\figref[B]{fig:3_biological_locomotion}). It does so via two pathways: one controlling speed by modulating the intrinsic period of RG oscillators, and one controlling gait by modulating the ``cross'' and ``diagonal'' RG connections that promote synchronization or alternation. \citep{Ausborn2019ComputationalModelingBrainstem}

\paragraph{Pattern Formation} The PF neural circuit within each limb converts half-center states into muscle activations (\figref[C]{fig:3_biological_locomotion}). This circuit of interneurons and motorneurons forms a 2-level hierarchy in which flexion and extension signals are expanded in dimensionality to produce precise activation signals for each muscle. \citep{Kim2022ContributionAfferentFeedback}

\paragraph{Afferent Feedback} The AF neural circuit within each limb uses sensory information to modulate RG and PF activity (\figref[C]{fig:3_biological_locomotion}). Muscle sensors produce length-, velocity-, and force-related signals, and foot sensors detect tactile stimulation. These signals trigger reflexes in PF and advance/delay half-center states in RG to entrain neural activity with musculoskeletal conditions. \citep{Kim2022ContributionAfferentFeedback}

\subsection{Architecture Units}
\label{sec:3_architecture_units}

\begin{figure}[t]
\centering
\includegraphics[width=\linewidth]{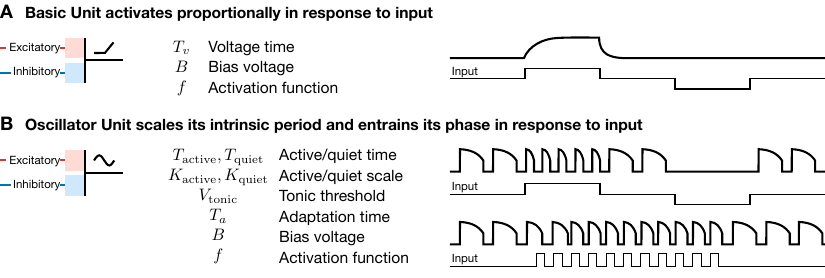}
\caption{
    \textbf{Architecture Units.} The architecture uses 2 neuron types, both of which output rate-coded activity and can receive excitatory or inhibitory synaptic input. For each type, we show the circuit schematic symbol (left), key hyperparameters (middle), and an example waveform response to input (right).
    \textbf{A}, The Basic unit is a typical neuron that activates proportionally in response to input once its internal voltage exceeds a threshold. This unit is used for most neurons in the architecture.
    \textbf{B}, The Oscillator unit is a special neuron that exhibits intrinsically bursting activity in the absence of inputs. It scales its intrinsic period in response to constant input, and it entrains its phase in response to periodic input. This unit is used for the flexor half-centers in the RG module.
}
\label{fig:3_architecture_units}
\end{figure}
Our NCAP architecture adopts a continuous-time framework for modeling neurons. We find that 2 computational units can capture the cell types in this circuit: a Basic unit and an Oscillator unit (\figref{fig:3_architecture_units}). Many neuromechanical modeling works, including \cite{Danner2017ComputationalModelingSpinal}, use biophysical neuron models that incorporate the conductances, reversal potentials, and activation/inactivation dynamics of ion channel currents (for instance, a persistent sodium current for bursting). Such complexity is not needed for AI purposes, so we follow \cite{Bhattasali2022NeuralCircuitArchitectural} by simplifying these neurons to create computational units with fewer and more interpretable hyperparameters. We describe below the main properties of these units. For details and equations, please see \autoref{sec:a1_architecture_units}.

\paragraph{Basic Unit} This neuron model is standard in computational neuroscience. Rate-coded inputs raise or lower the internal voltage, which is leaky. If the internal voltage rises beyond a threshold, the neuron generates rate-coded output activity according to an activation function.

\paragraph{Oscillator Unit} This neuron model abstracts an intrinsically bursting neuron \citep{Danner2017ComputationalModelingSpinal}. It generates oscillating output activity in the absence of inputs. In response to inputs, it scales its active and quiet phases, and it transitions to a silent mode under strong inhibition and to a tonic mode under strong excitation. These properties enable the unit to shift its oscillation phase to pulse waveforms, and entrain its oscillation phase to periodic waveforms. We provide a detailed derivation and evaluation of this model in a concurrent manuscript \oscref.

\subsection{Architecture Structure}
\label{sec:3_architecture_structure}

\begin{figure}[t]
\centering
\includegraphics[width=\linewidth]{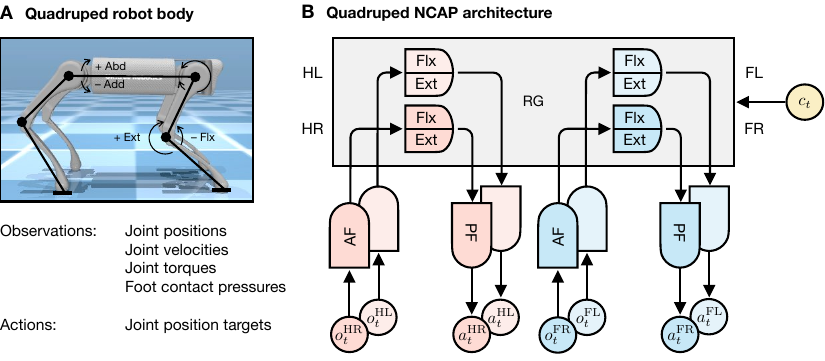}
\caption{
    \textbf{Architecture Structure.}
    \textbf{A}, The Quadruped robot has 4 limbs, each with 3 joints (hip, thigh, calf) and 1 foot pad. The ANN agent receives proprioceptive/pressure observations, and it produces joint position target actions that are converted to actuator torque commands by a low-level PD controller.
    \textbf{B}, The Quadruped NCAP architecture mirrors mammalian locomotion circuits. The RG module receives brainstem commands $c_t$ that set the speed and gait. Each limb has an AF module that uses leg observations $o_t$ to modulate RG oscillators, as well as a PF module that converts RG half-center states into leg actions $a_t$.
}
\label{fig:3_architecture_structure}
\end{figure}

\paragraph{Robot Body} We target a standard robotic body in order to investigate the effectiveness of architectural priors in AI settings (\figref[A]{fig:3_architecture_structure}). Unlike animals, the robot does not use highly redundant muscles that produce linear force; instead, it uses a single rotational motor per joint that produces torque. Our architecture controlling this body must translate between biological neural circuitry and the artificial agent interface. For the observation space, the agent receives joint positions, velocities, and torques as well as foot contact pressures. For the action space, the agent produces target joint positions, which are converted to actuator torque commands by a low-level PD controller (\autoref{sec:a2_simulated_robot}). While this interface is simplified compared to biology, it reasonably approximates how muscles have a net effect of setting a joint's equilibrium position and stiffness \citep{Shadmehr1993ControlEquilibriumPosition}.

\paragraph{Rhythm Generation} We adapt the RG circuit from \cite{Danner2017ComputationalModelingSpinal} to use our simplified Basic and Oscillator units. We eliminate redundant connections and introduce new nomenclature for the neuron types. As discovering the RG connections is not an aim of this work, we hand-tune the RG weights to produce the appropriate gait transitions in response to brainstem input (\autoref{fig:a1_gait_transitions}), using the reported connection strengths from \cite{Danner2017ComputationalModelingSpinal} as a starting point. We then freeze the weights during training. For a full diagram of the RG module, please see \autoref{sec:a1_rhythm_generation}.

\paragraph{Brainstem Command} We use a single brainstem command~$c_t$ to control the RG module, which is task-specific and frozen during training. This enables us to control the gait pattern expressed at a particular speed. If not frozen, the training usually prefers a bound gait, which is fastest and enables the agent to maximize reward even on slow speed tasks. Having an architectural prior thus affords us fine-grained control of the resulting behavior without additional reward/imitation priors.

\paragraph{Pattern Formation} We use a linear PF layer to map flexor/extensor half-center outputs to leg actions~$a_t$. We initialize PF weights with coarse magnitudes and correct signs (producing negative joint positions for flexion and positive joint positions for extension) that are constrained after each update. We share weights among forelimbs and hindlimbs to exploit bilateral symmetry (\figref{fig:a4_interpretability_ncap_untrained_compact}), and we apply the overparameterization trick (\autoref{sec:a1_overparameterization_trick}) to expand the dimensionality of weights during training and collapse it during testing.

\paragraph{Afferent Feedback} We use a linear AF layer to map leg observations~$o_t$ to flexor/extensor inputs. We normalize observations to the range [$-1, 1$], then rectify them into positive and negative components, since firing rates cannot be negative. We initialize certain AF weights with coarse magnitudes to encode the known effects of leg loading and position on half-centers, and we constrain their signs (\figref{fig:a4_interpretability_ncap_untrained_compact}). We initialize unknown AF weights at zero and leave them unconstrained. We also utilize bilateral sharing and the overparameterization trick.
\section{Experiments}
\label{sec:4_experiments}

\paragraph{Tasks} We train our architecture on simulated tasks built atop the MuJoCo physics engine \citep{Todorov2012MuJoCoPhysicsEngine} that control the Unitree A1 robot (\autoref{sec:a2_simulated_robot}). The tasks are structured as 15-second episodes during which the robot must locomote forwards at a fixed target speed of Walk (0.5 m/s) or Run (1.0 m/s), and across terrains of Flat or Bumpy (\autoref{sec:a2_task_structure}). The tasks provide at each timestep a reward proportional to the running speed, with maximum reward of 1 at the target speed in the forward direction  (\autoref{sec:a2_task_rewards}). This task structure and reward design is based directly on \citet{Smith2022WalkParkLearning} in order to facilitate comparison to existing work.

\paragraph{Baselines}
We compare against multilayered perceptrons (MLPs) of 2 hidden layers. By default, we compare to MLP(256,256), which is a reasonably sized architecture commonly used in the AI and robotics literature. Importantly, we choose to baseline against MLPs as they exemplify an architecture \textit{without priors}, which contrasts with our NCAP architecture \textit{with priors}, facilitating a clean comparison. Our goal in this work is not to compare how different classes of prior stack up generally, but rather to explore the value of neural circuit-inspired architectural priors in particular. As architectural priors are somewhat orthogonal to other forms of priors (including reward, training curricula, and imitation priors), future work could combinatorially combine our prior with others.

\paragraph{Algorithm}
We train both NCAP and MLP architectures using evolution strategies to maximize episodic return (\autoref{sec:a3_evolution_strategies}). Such gradient-free optimization is easiest to use with our NCAP architecture, and it has successfully and popularly been used to train MLPs in continuous control \citep{Salimans2017EvolutionStrategiesScalable}. In preliminary experiments, we compare evolution strategies to standard on-policy and off-policy reinforcement learning algorithms (\autoref{sec:a3_reinforcement_learning}), and we confirm similar performance across algorithms when training MLPs on our tasks (\autoref{sec:a4_performance_and_data_efficiency}).

\subsection{Performance and Data Efficiency}
\label{sec:4_performance_fixed_speed}

\begin{figure}[t]
\centering
\includegraphics[width=\linewidth]{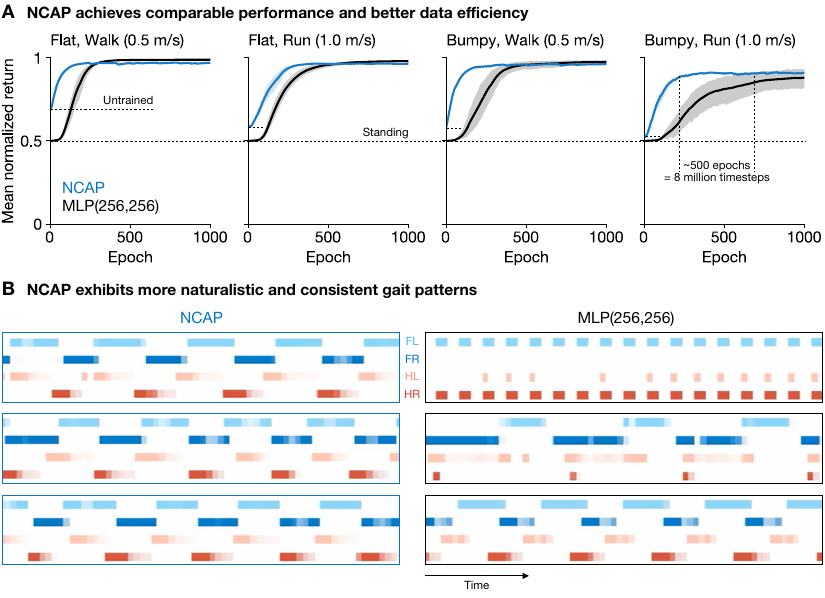}
\caption{
    \textbf{Performance and Data Efficiency.}
    \textbf{A}, Performance curves across tasks. Solid lines are mean normalized returns across 10 training seeds, each tested for 5 episodes per epoch. Shaded areas are 95\% bootstrapped confidence intervals. Maximum normalized episodic return is 1, and a policy that outputs all zeros achieves 0.5 by standing still.
    NCAP matches or exceeds the asymptotic performance of MLP, with superior data efficiency. NCAP also demonstrates better initial performance since it is an effective prior.
    \textbf{B}, Footfall plots across 3 training seeds on the Flat/Walk task. Colored segments encode the foot contact pressures during stance, while blank segments indicate the limb is in swing. NCAP exhibits qualitatively more naturalistic and consistent gaits than MLP, despite their quantitatively similar asymptotic performances.
}
\label{fig:4_performance_fixed_speed}
\end{figure}

NCAP successfully learns to locomote across various speeds and terrains (\figref[A]{fig:4_performance_fixed_speed}). For representative examples of NCAP's behavior and neural activity, please see \webref{Videos 1}.

How does NCAP compare to MLP on performance and data efficiency? NCAP achieves comparable asymptotic performance to MLP across tasks (\figref[A]{fig:4_performance_fixed_speed}). Moreover, due to its priors, an untrained NCAP achieves significantly better initial performance than an untrained MLP. The performance of NCAP improves with training as the AF and PF weights are refined. In addition, NCAP's training trajectories are less variable than the MLP's.

Interestingly, NCAP appears to train more data efficiently than MLP for the harder Bumpy tasks. For instance, on the Bumpy/Run task, NCAP reaches asymptotic performance about 500 epochs (or 8 million timesteps) before MLP. However, NCAP reaches slightly lower asymptotic performance than MLP for the easier Flat tasks. We attribute this to a regularization effect in NCAP, as it is constrained in the solutions it can learn. In contrast, MLP can learn to exploit the simulator for additional performance gains, which is easier to do on Flat than Bumpy tasks.

This is supported by the qualitative performance of NCAP and MLP. Using footfall plots, we examine learned gaits on the Flat/Walk task for different training seeds (\figref[B]{fig:4_performance_fixed_speed}, \webref{Videos 2}). MLP develops a good walking gait on seed 3, a mediocre limping gait on seed 2, and a failed on-the-floor shuffle on seed 1; this behavior is starkly evident in \webref{Videos 2}. Notably, this high variability is occluded in the performance curve (\figref[A]{fig:4_performance_fixed_speed}). In contrast, NCAP exhibits more naturalistic and consistent gaits due to its RG prior. Such gaits would require more priors to elicit from MLP (for example, reward or imitation priors).

\subsection{Parameter Efficiency}
\label{sec:4_parameter_efficiency}

\begin{figure}
\centering
\includegraphics[width=\linewidth]{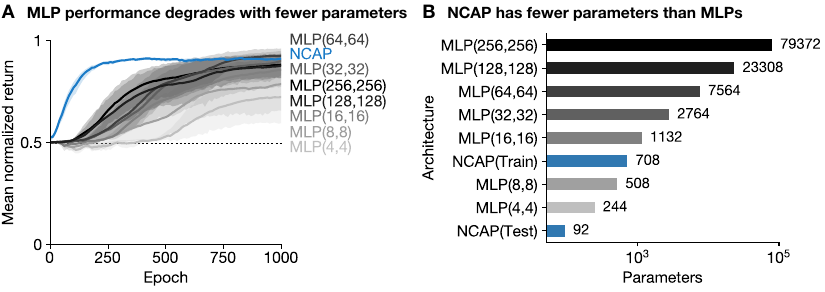}
\caption{
    \textbf{Parameter Efficiency.}
    \textbf{A}, Performance curves across MLP sizes on the Bumpy/Run task. Smaller MLPs achieve lower asymptotic performance and worse data efficiency. Therefore, having fewer parameters is insufficient to account for NCAP's advantages.
    \textbf{B}, Parameter count across architectures (log scale). NCAP(Test) has fewer parameters than MLP(4,4), and even NCAP(Train) with the overparameterization trick (\autoref{sec:a1_overparameterization_trick}) has orders of magnitude fewer parameters than MLPs at typical sizes.}
\label{fig:4_parameter_efficiency}
\end{figure}

Are the performance and data efficiency advantages of NCAP merely due to having fewer parameters? We test MLP with fewer parameters by varying the hidden layer sizes from 4 to 256. Surprisingly, performance and data efficiency degrade significantly (\figref[A]{fig:4_parameter_efficiency}), showing that it is not merely having fewer parameters that is beneficial. Rather, the specific structure of NCAP matters.

It is this structure that enables NCAP to perform well yet require dramatically fewer parameters than MLPs (\figref[B]{fig:4_parameter_efficiency}). During testing, NCAP uses only 92 parameters (\autoref{fig:a4_interpretability_ncap_trained_compact}), which is 3 orders of magnitude fewer than MLP(256,256) with 79,372 parameters (\autoref{fig:a4_interpretability_mlp_trained_full}), and it is fewer than even MLP(4,4). During training, NCAP with the overparameterization trick has 708 parameters (\autoref{fig:a4_interpretability_ncap_trained_full}), which is 2 orders of magnitude smaller than MLP(256,256), and it is comparable to MLP(16,16). Notably, many of these MLPs are much smaller than the sizes typically used in practice. Moreover, in settings like actor-critic reinforcement learning that often use similarly sized actor and critic networks as well as moving average copies of those networks, the number of required parameters can quadruple. NCAP's parameter efficiency could be particularly advantageous for deployment in resource-constrained environments, like a robot's onboard compute.

\subsection{Generalization to Terrain and Body Variations}
\label{sec:4_generalization_simulation}

\begin{figure}
\centering
\includegraphics[width=\linewidth]{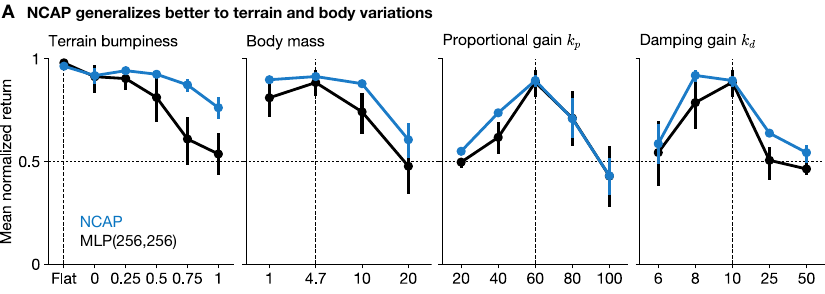}
\caption{
    \textbf{Generalization to Terrain and Body Variations.}
    \textbf{A}, Performance on unseen terrain and body variations. Architectures are trained in the condition indicated with a vertical dashed line, then tested in altered conditions. The variations included terrain bumpiness (parameterized from 0 to 1; \autoref{fig:a2_terrain_bumpiness}), body mass (in kilograms), or proportional/damping gains of the low-level PD controller. NCAP's generalization matches, and often exceeds, that of MLP.
}
\label{fig:4_generalization_simulation}
\end{figure}

How well does NCAP generalize to unseen environments compared to MLP? We evaluate the architectures across a variety of terrain and body variations (\figref{fig:4_generalization_simulation}). In each setting, the architectures are trained in one condition, then tested in altered conditions. Across these variations, NCAP's generalization matches, and often exceeds, that of MLP. Surprisingly, MLP performance degrades dramatically on Bumpy terrain, despite the differences in bumpiness seeming minor by human standards (\figref{fig:a2_terrain_bumpiness}). In contrast, NCAP performs more robustly.

\subsection{Generalization to the Physical Robot}
\label{sec:4_generalization_physical}

\begin{figure}
\centering
\includegraphics[width=\linewidth]{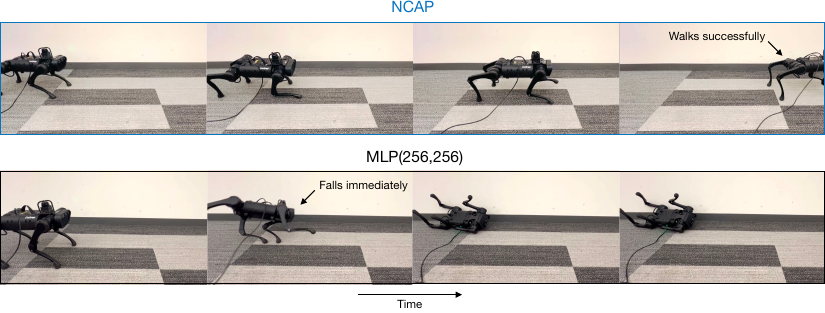}
\caption{
    \textbf{Generalization to the Physical Robot.}
    Video frames (\webref{Videos 3}) of a representative deployment trial on the physical robot. NCAP walks successfully despite the large domain gap, while MLP falls immediately due to its erratic and unstable actions.
}
\label{fig:4_generalization_physical}
\end{figure}

How well does NCAP generalize to the real world compared to MLP? We deploy the architectures to a physical Unitree A1 quadruped robot after training on the Bumpy/Walk task (\autoref{sec:a2_physical_robot}). We expect the domain gap to be large since we do not perform controller tuning or system identification with the physical robot, the simulated training does not apply aggressive domain randomization (a form of task prior), and the architectures do not have a mechanism for online adaptation.

MLP falls immediately due to its erratic and unstable actions, often launching the robot aggressively into the wall (\figref{fig:4_generalization_physical}). Despite our best efforts, we cannot elicit a successful walking trial. In contrast, NCAP is remarkably robust to the large domain gap, walking successfully on most trials, though with less smoothness than in simulation (\figref{fig:4_generalization_physical}). We attribute NCAP's success to a combination of the RG module maintaining a stable rhythm in the face of sensor noise and the AF module triggering corrective responses in the face of perturbations. Lastly, we deploy an untrained NCAP and observe that it is stable and produces slight walking movements with small foot displacements (\webref{Videos 3}).

\section{Discussion}
\label{sec:5_discussion}

In this work, we introduce Quadruped NCAP, a biologically inspired ANN architecture for quadruped locomotion based on neural circuits in the limbs and spinal cord of mammals. Our architecture achieves good initial performance and comparable final performance to MLPs, while using less data and orders of magnitude fewer parameters. Our architecture also exhibits better generalization to task variations, even admitting deployment on a physical robot without standard sim-to-real methods. 

\paragraph{Limitations} Our study faces several limitations. First, we rely on a hand-tuned RG module and BC command, which might not be the optimal parameters that a learning-based approach could discover. Second, we neglect musculoskeletal factors in the quadruped action space that could make learning easier or more robust. 

\paragraph{Future Work}
We focus on fixed speed locomotion in this work, but an obvious next step is to enable the RG to transition between gaits in a speed-dependent manner, which may largely involve a higher-level controller that alters brainstem commands for different speeds. Another extension is to add postural adjustment, turning, and righting mechanisms to the architecture based on understanding of the underlying neural circuits. Finally, as architectural priors are somewhat orthogonal to other forms of priors, it may be beneficial to train NCAP with additional reward, task, or imitation priors.

Overall, we believe that this work shows that neural circuits can provide valuable architectural priors for locomotion in more complex animals and encourages future work in yet more complex sensorimotor skills.

\subsubsection*{Acknowledgments}
Thanks for insightful discussions to Pavel Tolmachev, Liam McCarty, Anthony Chen, Ulyana Piterbarg, Jeff Cui, Ben Evans, Siddhant Haldar, Noah Amsel, Yann LeCun, Tony Zador, and many others. This work was supported by the Fannie and John Hertz Foundation Fellowship (NXB); NSF award 2339096 (LP); ONR awards N00014-21-1-2758 and N00014-22-1-2773 (LP); and the Packard Fellowship for Science and Engineering (LP).

\subsubsection*{Author Contributions}
Conceptualization (NXB; LP, GWL). Investigation (NXB). Software: simulation (NXB), deployment (VP). Supervision (LP, GWL). Visualization (NXB). Writing: original draft (NXB, VP), review and editing (NXB, VP, LP, GWL).

\bibliographystyle{template/iclr2025_conference}
\bibliography{bibliography}
\FloatBarrier
\newpage

\appendix
\counterwithin{figure}{section}
\section{Architecture Details}
\label{sec:a1}

\subsection{Architecture Units}
\label{sec:a1_architecture_units}

\paragraph{Basic Unit} The Basic unit is a typical rate-coded neuron with continuous dynamics, which integrates weighted input signals to generate a target $x$ for the internal voltage $v$.
\begin{gather*}
    \text{Hyperparameters: } T_v, B, f \\
    x = \text{clip}( B + \sum\nolimits_i{w_i y_i}, -1, 1 ) \\
    \frac{T_v}{4} \frac{dv}{dt} = x - v
\end{gather*}

The neuron’s activation function $f$ determines the voltage-firing relationship:
\begin{align*}
    &y = f(v),\quad f = \text{clip}(v, 0, 1)
\end{align*}

\paragraph{Oscillator Unit} The Oscillator unit is a relaxation oscillator with discrete-continuous dynamics on 2 variables: a discrete internal voltage $v \in {-1, +1}$ and a continuous adaptation variable $a \in [0, 1]$.
\begin{gather*}
    \text{Hyperparameters: } T_a, T_\text{active}, T_\text{quiet}, K_\text{active}, K_\text{quiet}, V_\text{tonic}, B, f \\
    T'_\text{active} = \frac{4T_\text{active}}{T_a},\quad
    T'_\text{quiet} = \frac{4T_\text{quiet}}{T_a}
\end{gather*}

The adaptation thresholds $a^\text{0, 1}_\text{quiet, active}$ are calculated that determine the min/max values of adaptation $a$ at which the internal voltage $v$ jumps between active ($+1$) and quiet ($-1$) states:
\begin{align*}
    a_\text{active}^0 &= \frac{1 - \exp(T'_\text{quiet})}{1 - \exp(T'_\text{active} + T'_\text{quiet})} \\
    a_\text{quiet}^0 &= a_\text{active}^0 \cdot \exp(T'_\text{active}) \\
    a_\text{active}^1 &= \frac{1 - \exp(T'_\text{quiet} \cdot K_\text{quiet})}{1 - \exp(T'_\text{active} \cdot K_\text{active} + T'_\text{quiet} \cdot K_\text{quiet})} \\
    a_\text{quiet}^1 &= a_\text{active}^1 \cdot \exp(T'_\text{active} \cdot K_\text{active})
\end{align*}

Depending on the strength of the input signal at a given time, the quiet-to-active and active-to-quiet adaptation thresholds interpolate between the calculated min/max values. Adaptation $a$ exponentially decays towards 0 when the neuron is active (the neuron depletes the adaptation variable) and towards 1 when quiet (the neuron replenishes the adaptation variable). Voltage $v$ jumps instantaneously to a new state when an adaptation threshold is reached.
\begin{gather*}
    x = \text{clip}( B + \sum\nolimits_i{w_i y_i}, -1, 1 ) \\
    z = \text{clip}( x, 0, 1) \\
    a_\text{active} = \text{interpolate}( z, a_\text{active}^0, a_\text{active}^1 ) \\
    a_\text{quiet} = \text{interpolate}( z, a_\text{quiet}^0, a_\text{quiet}^1 ) \\
    \frac{T_a}{4} \frac{da}{dt} = \left\{\begin{array}{lll}0-a &\text{if } v = +1 &\text{(active)}\\1-a &\text{if } v = -1 &\text{(quiet)}\end{array}\right. \\
    v^{(t + dt)} := \left\{\begin{array}{lll}
        -1 & \textrm{if } a^{(t)} \leq a_\text{active}^{(t)} \text{ and } x^{(t)} \leq V_\text{tonic} & \text{(active $\rightarrow$ quiet)}\\
        +1 & \textrm{if } a^{(t)} \geq a_\text{quiet}^{(t)} \text{ and } x^{(t)} \geq 0 & \text{(quiet $\rightarrow$ active)}\\
        v^{(t)} & \text{otherwise} &
        \end{array}\right.
\end{gather*}

The neuron’s activation function $f$ determines the voltage-adaptation-firing relationship:
\begin{align*}
    &y = f(v, a, x),\quad f = \left\{\begin{array}{lll}
    \text{interpolate}(a, 0.5, 1.0) &\text{if } v = +1 &\text{(active)}\\
    0 &\text{if } v = -1 &\text{(quiet)}\end{array}\right.
\end{align*}

We provide a detailed derivation and evaluation of this model in a concurrent manuscript \oscref.

\subsection{Rhythm Generation (RG) Module}
\label{sec:a1_rhythm_generation}

\begin{figure}[ht]
\centering
\includegraphics[width=\linewidth]{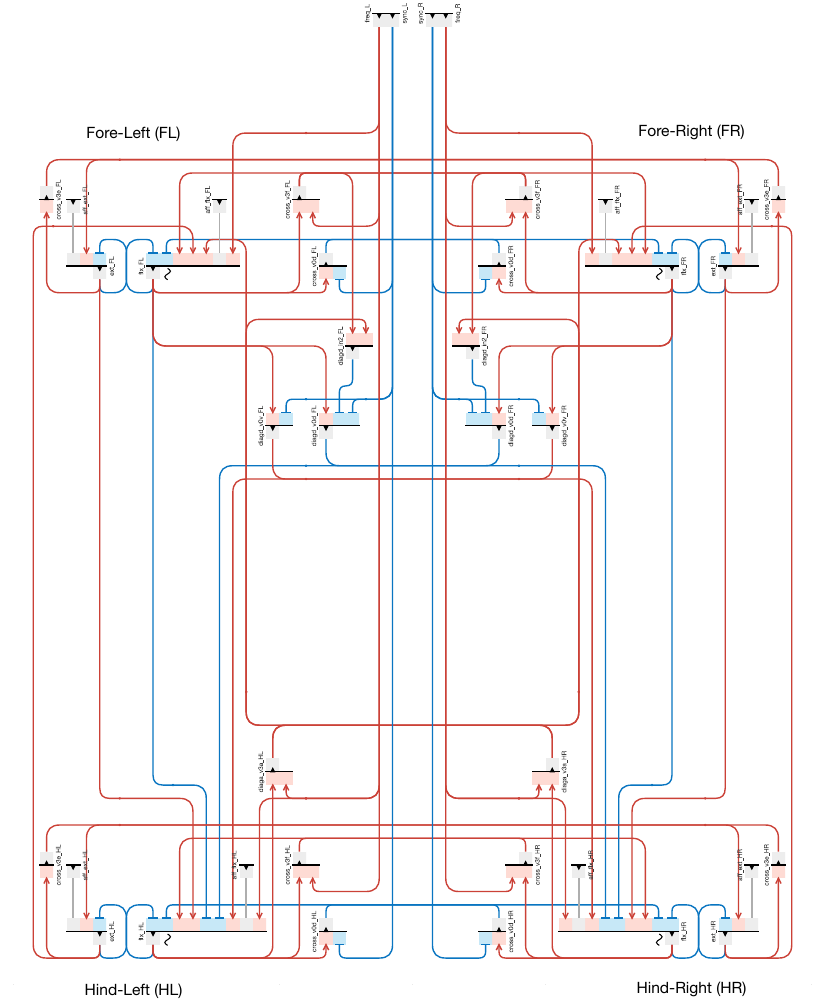}
\caption{
    \textbf{Rhythm Generation, Full.} Neural circuits for rhythm generation (RG) and brainstem command (BC) at cell-type-specific resolution, adapted from \citet{Danner2017ComputationalModelingSpinal}. For a progressive breakdown, please see \autoref{fig:a1_rhythm_generation_detailed}.
}
\label{fig:a1_rhythm_generation_full}
\end{figure}

\begin{figure}[ht]
\centering
\includegraphics[width=\linewidth]{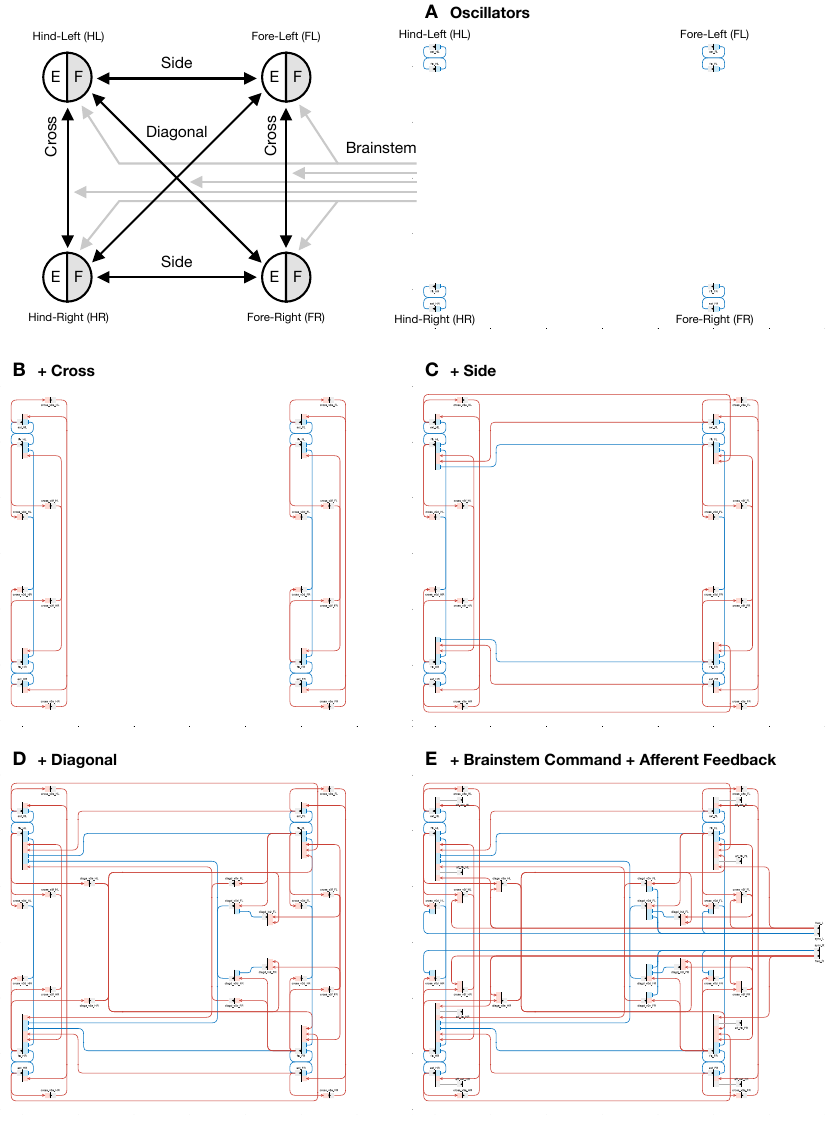}
\caption{
    \textbf{Rhythm Generation, Breakdown.} Neural circuits for rhythm generation (RG) and brainstem command (BC) at cell-type-specific resolution, adapted from \citet{Danner2017ComputationalModelingSpinal}. Each limb is controlled by \textbf{(A)} reciprocally inhibiting half-centers consisting of a flexor Oscillator unit and an extensor Basic unit. Between limbs, half-centers communicate through \textbf{(B)} cross, \textbf{(C)} side, and \textbf{(D)} diagonal connections, and they are modulated by \textbf{(E)} brainstem command and afferent feedback.
}
\label{fig:a1_rhythm_generation_detailed}
\end{figure}

\begin{figure}[t]
\centering
\includegraphics[width=\linewidth]{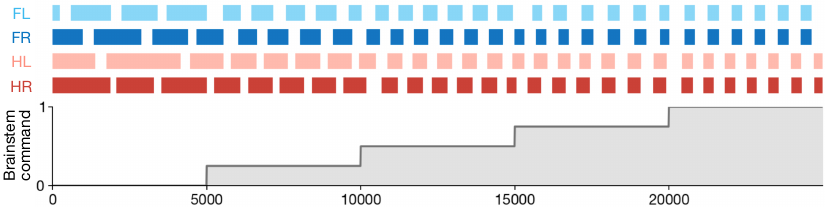}
\caption{
    \textbf{Rhythm Generation, Gait Transitions.} Predicted footfall plots from RG extensor half-center state. The rhythm generation circuit transitions between gaits (walk → trot → bound) as the brainstem command increases from 0 to 1. This enables a higher-level controller to modulate the gait for variable speeds.
}
\label{fig:a1_gait_transitions}
\end{figure}

The RG circuit diagram can be visualized in full 
(\autoref{fig:a1_rhythm_generation_full}) or through a progressive breakdown (\autoref{fig:a1_rhythm_generation_detailed}). The RG weights are tuned to produce the appropriate gait transitions in response to increasing brainstem command (\autoref{fig:a1_gait_transitions}).

\FloatBarrier

\subsection{Overparameterization Trick}
\label{sec:a1_overparameterization_trick}

We discover an important technique to improve the trainability of our architecture, which we call ``the overparameterization trick''. 

Consider a linear layer $\vy = \mW \vx$. During training, the weight matrix $\mW$ can be artificially expanded in dimensionality along the row dimension and/or column dimension to produce a larger weight matrix $\mW'$. The former produces a correspondingly larger output vector $\vy'$, while the latter necessitates a correspondingly larger input vector $\vx'$. The larger output $\vy'$ can be transformed into the original output $\vy$ by summing expanded elements. The original input $\vx$ can be transformed into the larger input $\vx'$ by copying and concatenating elements. Thus, the expanded layer maintains the original input and output dimensions, but with a larger weight matrix and new hardcoded expansion and compression operations at the interface. We find empirically that such overparameterization improved learning, presumably due to special dynamics of traversing high-dimensional loss landscapes with our point-based evolutionary strategies algorithm (\autoref{sec:a3_evolution_strategies}), but better theoretical understanding of this phenomenon is needed.

During testing, the overparameterized weight matrix $\mW'$ can be collapsed by summing along the expanded rows and/or columns to produce a smaller matrix with the size of the original $\mW$. Mathematically, this does not change the computation, as it exploits the linearity of matrix multiplication.

For a simple example, consider the case of 1D inputs and outputs.

If the weight is expanded row-wise:
$$
y = wx
\quad\underset{\textrm{expand}}{\longrightarrow}\quad
y =
\sum_{\text{rows}}
\begin{bmatrix}
w'_1 \\ w'_2 
\end{bmatrix}
x
= (w'_1 + w'_2) x
\quad\underset{\textrm{compress}}{\longrightarrow}\quad
y = wx
$$

If the weight is expanded column-wise:
$$
y = wx
\quad\underset{\textrm{expand}}{\longrightarrow}\quad
y =
\begin{bmatrix}
w'_1 & w'_2 
\end{bmatrix}
\begin{bmatrix}
x \\ x
\end{bmatrix}
= (w'_1 + w'_2) x
\quad\underset{\textrm{compress}}{\longrightarrow}\quad
y = wx
$$

\begin{figure}[!ht]
\centering
\includegraphics[width=\linewidth]{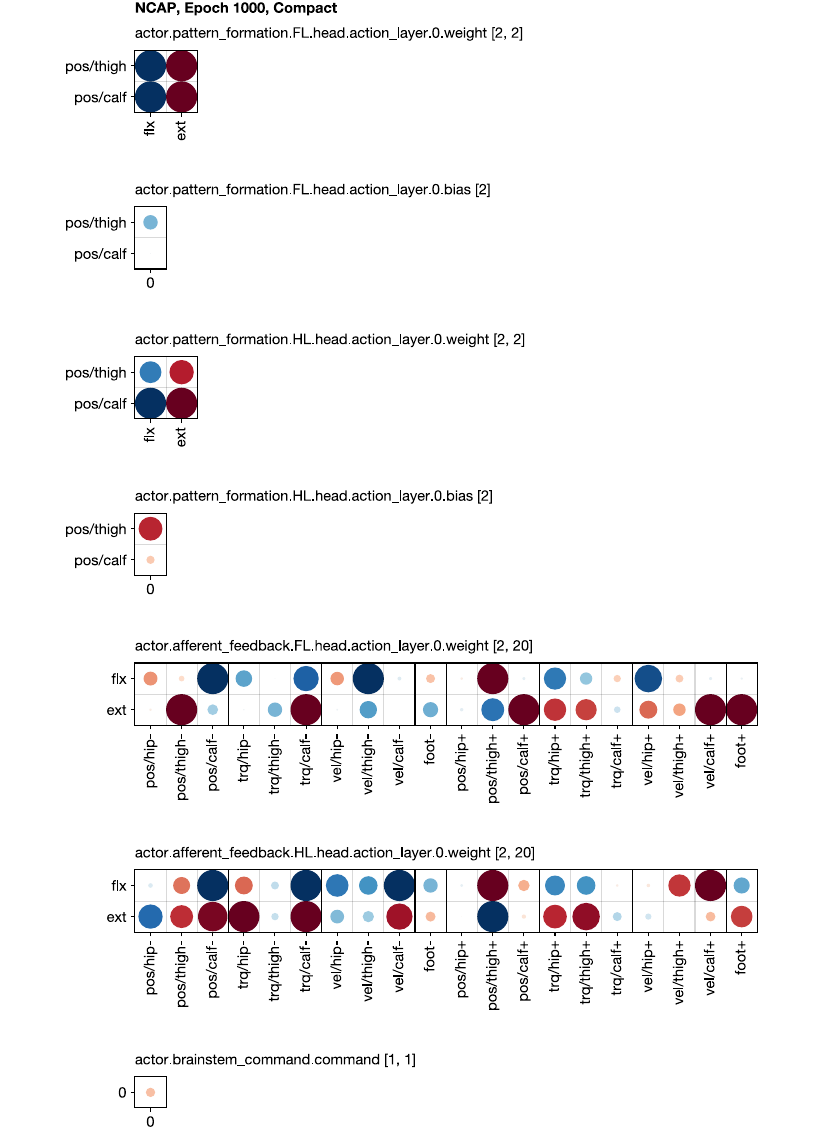}
\caption{
    \textbf{Interpretability, NCAP Architecture, Trained, Compact.}
    Weights plot of the trained NCAP in its compact/testing variant.
}
\label{fig:a4_interpretability_ncap_trained_compact}
\end{figure}

\begin{figure}[!ht]
\centering
\includegraphics[width=0.6\linewidth]{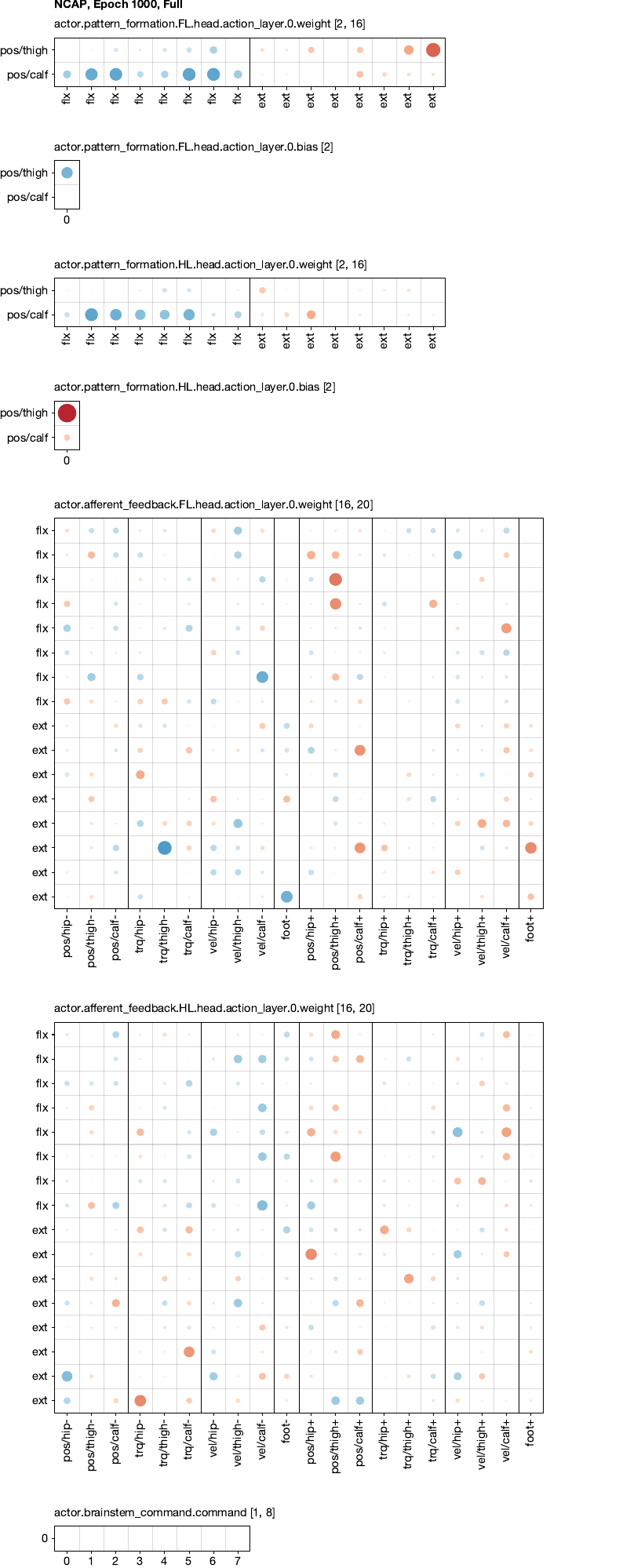}
\caption{
    \textbf{Interpretability, NCAP Architecture, Trained, Full.}
    Weights plot of the trained NCAP in its overparameterized/training variant.
}
\label{fig:a4_interpretability_ncap_trained_full}
\end{figure}

For this work, the technique enables NCAP to leverage the benefits of training with overparameterized networks using an expanded architecture (\figref{fig:a4_interpretability_ncap_trained_full}), which is compressed for testing  (\figref{fig:a4_interpretability_ncap_trained_compact}).
\newpage
\section{Environment Details}
\label{sec:a2}

\subsection{Simulated Robot}
\label{sec:a2_simulated_robot}

We train our architecture on simulated tasks built using DeepMind Composer \citep{Tassa2020DmControlSoftware} atop the MuJoCo physics engine \citep{Todorov2012MuJoCoPhysicsEngine} using a Unitree A1 robot model imported from MuJoCo Menagerie \citep{Zakka2022MuJoCoMenagerieCollection}.

The observation and action interfaces with the agent are normalized to [$-1, 1$]. The action space is designed with a default standing pose corresponding to actions of 0 and joint limits corresponding to actions of $\pm 1$.

A proportional-derivative (PD) controller is employed to convert the target joint positions generated by the agent $\vq_\text{target} = \va_t$ into joint torque commands for the actuators:
\[
\vtau = k_p (\vq_\text{target} - \vq) - k_d \cdot \dot{\vq}
\]
with joint positions $\vq$, joint velocities $\dot{\vq}$, proportional gain $k_p$, and derivative gain $k_d$.

The simulation runs with a control timestep of 0.03 seconds and a physics timestep of 0.001 seconds.

\subsection{Task Structure}
\label{sec:a2_task_structure}

\begin{figure}[ht]
\centering
\includegraphics[width=\linewidth]{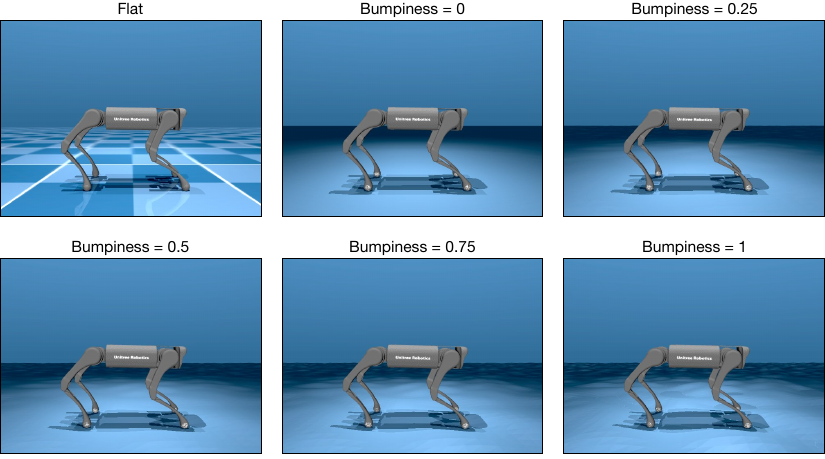}
\caption{
    \textbf{Terrain Bumpiness.} The Flat and Bumpy terrains have different frictions. The Bumpy terrain allows bumpiness to be parameterized between 0 and 1.
}
\label{fig:a2_terrain_bumpiness}
\end{figure}

The tasks are formulated as 15-second episodes. The agent incurs a penalty and the episode resets if the robot falls over or if any part of its base touches the ground. At the start of each episode, the friction coefficient of the foot is randomized. The agent is trained to move at a fixed velocity of Walk (0.5 m/s) or Run (1.0 m/s) over either Flat or Bumpy terrain (\autoref{fig:a2_terrain_bumpiness}).

\subsection{Task Rewards}
\label{sec:a2_task_rewards}

We adopt the reward function from \citet{Smith2022WalkParkLearning}, which uses the forward linear velocity in the robot frame $v_x$, the target forward velocity $v_x^{\text{target}}$, and the angular yaw velocity $\omega_z$.

The overall reward combines a forward velocity reward and a rotational penalty. The tolerance function $r_v$ encourages the agent to maintain a forward velocity near the target with some allowable deviation, while the rotational penalty reduces unnecessary rotational movements, ensuring stable and forward-directed motion:
\begin{gather*}
r(s, a) = r_v(s, a) - 0.1 \omega_z^2 \\
r_v(s, a) = \left\{\begin{array}{ll}
    1 & v_x \in [v_t, 2v_t] \\
    0 & v_x \in (-\infty, -v_t] \cup [4v_t, \infty) \\
    1 - \frac{|v_x - v_t|}{2v_t} & \text{otherwise}
    \end{array}\right.
\end{gather*}

\FloatBarrier

\subsection{Physical Robot}
\label{sec:a2_physical_robot}

\paragraph{Deployment Setup}
The experiments are conducted on the Unitree A1 quadruped robot with 12 degrees of freedom (3 per leg). The robot has sensors for joint positions, joint torques, joint velocities, and foot contact pressures, and it has actuators that produce commanded torques at each joint. The communication between the robot and the control policy is performed through Lightweight Communications and Marshalling (LCM) \citep{Huang2010LCMLightweightCommunications} and a pybind build of \texttt{unitree\_legged\_sdk}. The control policy is deployed on the robot from a local workstation through a LAN connection, in order to ensure low-latency and high-reliability communication. 

\paragraph{Experimental Protocol} The deployment process includes an initial calibration phase to verify joint offsets and zero torque sensors, ensuring accurate sensor readings and actuator control. Notably, no system identification is performed with our simulated robot model. Despite substantial efforts to tune the MLP policy, including adjustments to proportional/derivative gains and initial placement, it struggles to generalize effectively to the physical robot. In contrast, our NCAP policy exhibits surprising robustness.

\newpage
\section{Algorithm Details}
\label{sec:a3}

\subsection{Evolution Strategies} 
\label{sec:a3_evolution_strategies}

\paragraph{Augmented Random Search (ARS)} ARS is an evolutionary strategies algorithm that spawns offspring networks by randomly perturbing the parent network's weights, scores offspring according to a fitness function, and selects the next parent by taking a fitness-weighted average of offspring weights. It can be approximately viewed as estimating gradients of the fitness function. \citep{Mania2018SimpleRandomSearch} 

\begin{table}[h!]
\centering
\begin{tabular}{@{}lll@{}}
\toprule
\textbf{Hyperparameter} & \textbf{ARS} \\ \midrule
Population size         & 256            \\
Mutation scale          & 0.1            \\
Number of workers       & 32             \\ \bottomrule
\end{tabular}
\caption{\textbf{Hyperparameters for ES algorithm.} Further details in our provided code.}
\end{table}

\subsection{Reinforcement Learning}
\label{sec:a3_reinforcement_learning}

\paragraph{Proximal Policy Optimization (PPO)} PPO is an on-policy reinforcement learning algorithm known for its stability. By maintaining a proximity constraint between the new and old policies, it balances exploration and exploitation, making it a reliable choice for continuous control tasks. \citep{Schulman2017ProximalPolicyOptimization}

\paragraph{Distributed Distributional Deep Deterministic Policy Gradient (D4PG)} D4PG is an off-policy reinforcement learning algorithm known for its data efficiency. By incorporating parallel rollouts and a distributional value function, it efficiently reuses gathered experience to learn continuous control tasks. \citep{Barth-Maron2018DistributedDistributionalDeterministic}

\begin{table}[h!]
\centering
\begin{tabular}{@{}lll@{}}
\toprule
\textbf{Hyperparameter}               & \textbf{D4PG}                           & \textbf{PPO}                            \\ \midrule
Actor network layers                  & (256, 256)                              & (64, 64)                                \\
Actor network activation function     & ReLU                                    & Tanh                                    \\
Critic network layers                 & (256, 256)                              & (64, 64)                                \\
Critic network activation function    & ReLU                                    & Tanh                                    \\
Observation normalizer                & Mean-Std                                & Mean-Std                                \\ \bottomrule
\end{tabular}
\caption{\textbf{Hyperparameters for D4PG and PPO algorithms.} Further details in the algorithm implementations from \citet{Pardo2021TonicDeepReinforcement}. }
\end{table}

\subsection{Computational Resources}
\label{sec:a3_computational_resources}
Training is performed on a high-performance computing cluster running the Linux Ubuntu operating system. The ES algorithm is parallelized over 32 cores. The RL algorithms are parallelized over 16 cores as minimal speedups are observed beyond that on our tasks; this kind of nonlinear scaling is consistent with reported performance tests in other work \citep{Salimans2017EvolutionStrategiesScalable}.
\newpage
\section{Supplemental Experiments}
\label{sec:a4}

\subsection{Performance and Data Efficiency}
\label{sec:a4_performance_and_data_efficiency}

\begin{figure}[h]
\centering
\includegraphics[width=\linewidth]{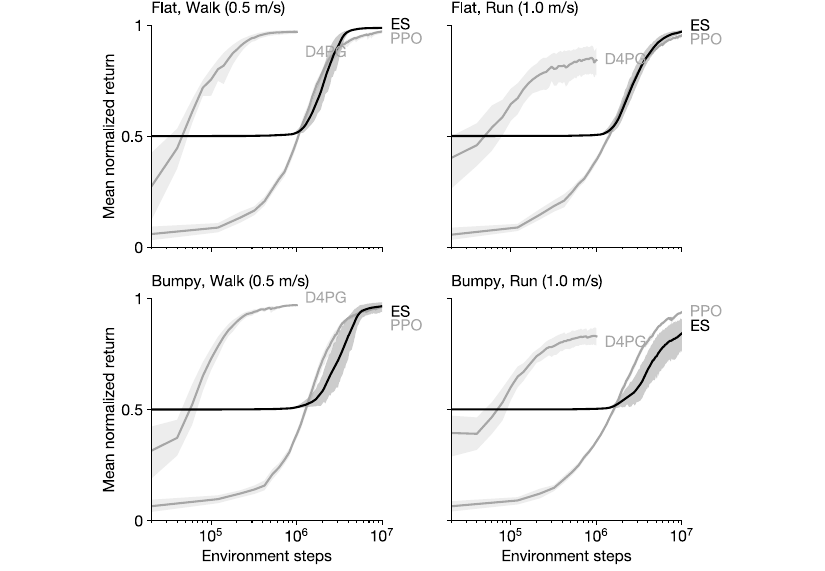}
\caption{
    \textbf{Performance and Data Efficiency, MLP Architecture, Various Algorithms.}
    erformance curves across various tasks. Solid lines are mean normalized returns across 10 training seeds, each tested for 5 episodes per epoch. Shaded areas are 95\% bootstrapped confidence intervals. The $x$-axis is environment steps to ensure a fair data efficiency comparison between algorithms, although the time efficiency between algorithms is significantly different. ES achieves comparable performance to PPO and D4PG on our tasks.
}
\end{figure}

\FloatBarrier

\subsection{Parameter Efficiency}
\label{sec:a4_parameter_efficiency}

\begin{figure}[h]
\centering
\includegraphics[width=\linewidth]{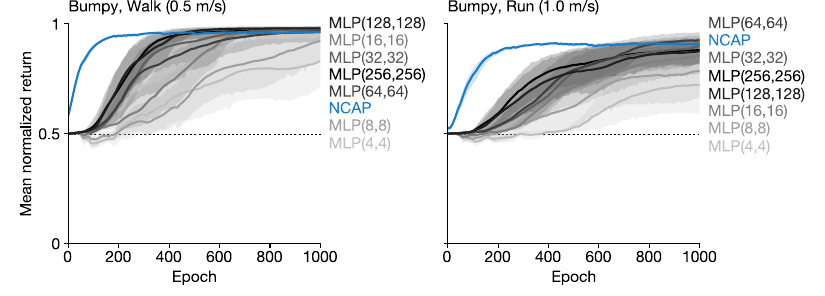}
\caption{
    \textbf{Parameter Efficiency, NCAP/MLP Architectures, Various Tasks.}
    Performance curves across MLP sizes on Bumpy tasks.  Smaller MLPs achieve lower asymptotic performance and worse data efficiency, which is more extreme in the harder Bumpy/Run task.
}
\end{figure}

\FloatBarrier

\subsection{Interpretability}
\label{sec:a4_interpretability}

\begin{figure}[!ht]
\centering
\includegraphics[width=\linewidth]{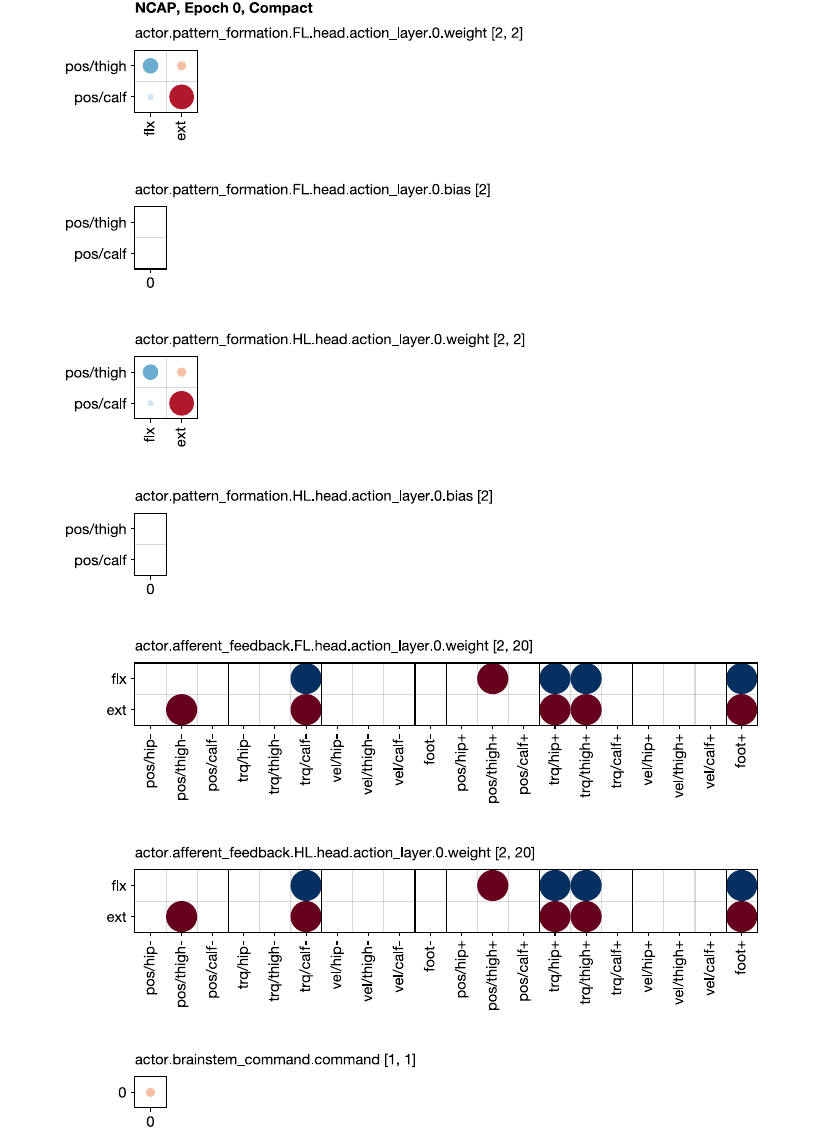}
\caption{
    \textbf{Interpretability, NCAP Architecture, Untrained, Compact.}
   Weights plot of the untrained NCAP. Weight sign is encoded in color, with excitatory (positive) weights as red and  inhibitory (negative) weights as blue. Weight magnitude is encoded in circle lightness and diameter. The non-zero elements of NCAP's weights are initialized with coarse magnitudes and constrained signs.
}
\label{fig:a4_interpretability_ncap_untrained_compact}
\end{figure}

\begin{figure}[!ht]
\centering
\includegraphics[width=0.85\linewidth]{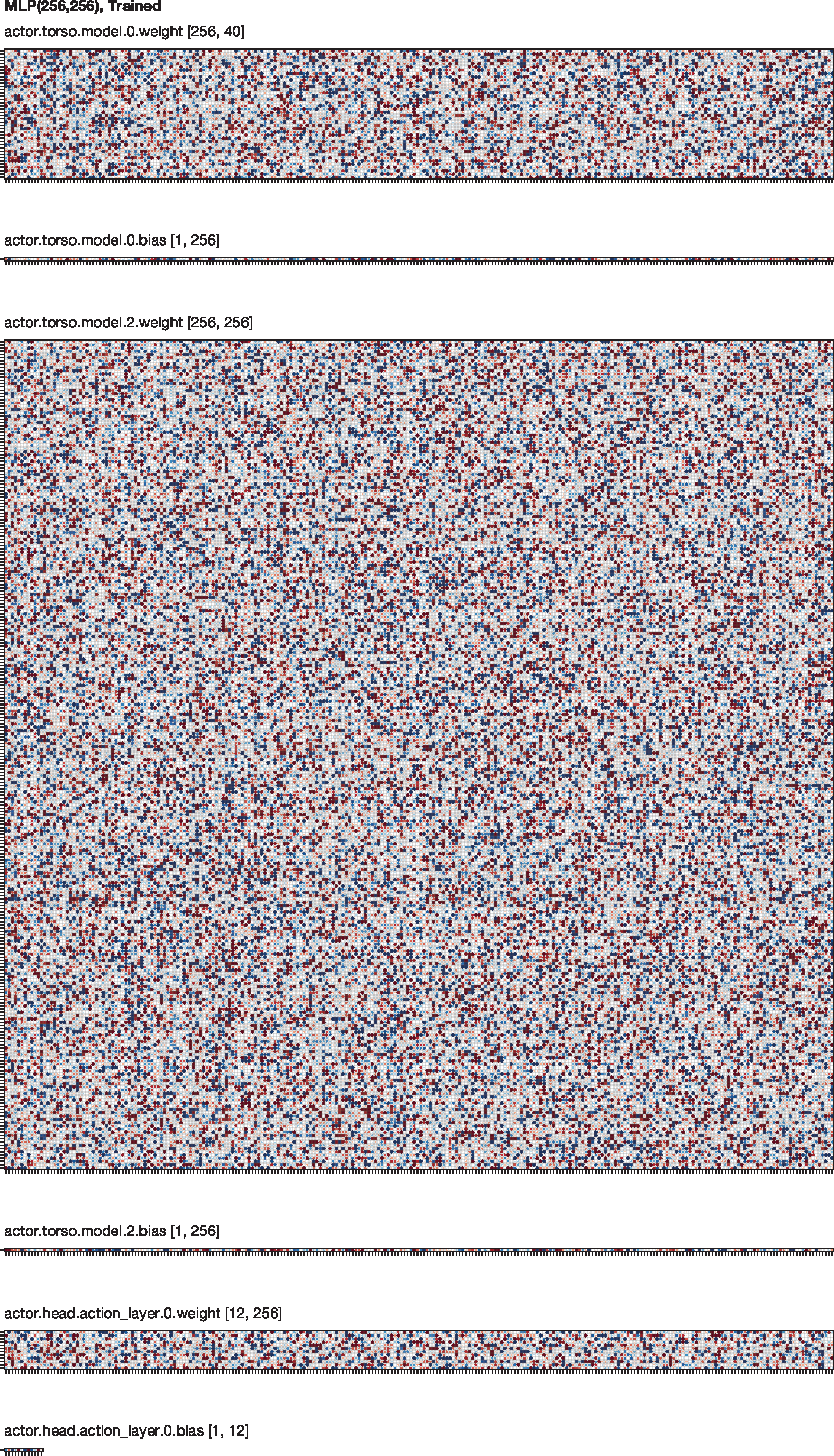}
\caption{
    \textbf{Interpretability, MLP Architecture, Trained, Full.}
    Weights plot of the trained MLP. Such weights are difficult to interpret, as the rows and columns of hidden weights lack fixed meanings, and weight signs can change freely during training.
}
\label{fig:a4_interpretability_mlp_trained_full}
\end{figure}

\FloatBarrier

\end{document}